\documentclass[sigplan,10pt]{acmart}
\acmConference[]{}
\acmISBN{} 
\acmDOI{} 
\startPage{1}

\setcopyright{none}

\usepackage{booktabs}   
\usepackage{subcaption} 
\usepackage{algorithmic}
\usepackage{graphicx}
\usepackage{multirow}
\usepackage{algorithm2e}
\usepackage{hyperref}
\usepackage{arydshln}
\usepackage{booktabs}
\usepackage{lineno}

\usepackage{cleveref}

\begin{document}

\title[Short Title]{A method of using RSVD in residual calculation of LowBit GEMM}         


\author{Hongyaoxing Gu}
\affiliation{
  \institution{lnstitute of Software Chinese Academy of Sciences}            
  \city{Beijing}
  \country{China}                    
}
\email{guhongyaoxing23@mails.ucas.ac.cn}          

\begin{abstract}
\quad The advancements of hardware technology in recent years has brought many possibilities for low-precision applications. However, the use of low precision can introduce significant computational errors, posing a considerable challenge to maintaining the computational accuracy.

We propose low-rank residuals quantized matrix multiplication(LRQMM) method which introduces low-rank approximation in residual compensation for dense low precision quantization matrix multiplication. It can bring several times accuracy improvement with only BLAS-2 level extra time overhead. Moreover, LRQMM is a completely data-free quantization method that does not require additional data for pre-training. And it only works with low precision GEMM operator, which is easy to couple with other methods.

Through experimentation, LRQMM can reduce the error of direct quantized matrix multiplication by \textbf{1-2} orders of magnitude, when dealing with larger matrix sizes, the computational speed is only reduced by approximately 20\%. In deep learning networks, LRQMM-4bit achieves 61.8\% ImageNet Top-1 accuracy in Resnet-50, while the Direct Quant accuracy is only 8.3\%. 
\end{abstract}

\ccsdesc[500]{Software and its engineering~General programming languages}
\ccsdesc[300]{Social and professional topics~History of programming languages}

\keywords{quantization, dense matrix multiplication, low-rank approximate}  

\maketitle

\section{Introduction}
\quad Dense matrix-matrix multiplication is a core function widely used in the fields of scientific and engineering computation as well as artificial intelligence. It is indispensable in solving mathematical problems such as linear equations \cite{lanczos1952solution}, eigenvalue problems \cite{abdi2007singular}, and matrix factorization \cite{li2005overview}, and also plays a key role in modern technological fields such as computer graphics \cite{he2016deep} and machine learning \cite{simonyan2014very}.

Traditional high-precision full-size matrix multiplication often cannot leverage specialized tensor acceleration hardware and can result in excessively large model sizes. To accelerate computation speed and reduce model size, approximate matrix multiplication algorithms have become a hot topic of research in recent years. This includes low-rank approximations, low-precision computations, and hash-based lookup techniques.

Low-precision computing is a technique in the field of computer science and engineering that improves computational efficiency. By using data types with lower bit widths, low-precision matrix multiplication can significantly reduce storage requirements and increase computational speed while reducing power consumption. This is particularly valuable in resource-constrained environments such as mobile devices and embedded systems, as well as in the field of large models \cite{jacob2018quantization,park2022lut}. With the support of new generation computing devices such as GPUs \cite{choquette2021nvidia}, NPUs \cite{hickmann2020intel,boutros2020beyond}, and TPUs \cite{jouppi2017datacenter}, Low-precision computations can offer significant speedup benefits. In the domain of numerical computation, low-precision operations are often utilized in iterative algorithms. The GMRES-IR algorithm \cite{higham2021exploiting,higham2019squeezing} computes the Cholesky decomposition at a lower precision and uses these factors as a preconditioning step for GMRES-based iterative refinement.

However, low-precision computing also brings challenges, such as precision loss and algorithm adaptability issues. To address this problem, researchers have developed a series of solutions, which include various quantization methods such as QAT \cite{liu2023llm}, LQ-Nets \cite{zhang2018lq}. However, these different methods all have certain limitations, requiring pre-training based on input data. But due to privacy and security concerns, the input data set may not be accessible. Besides, these quantization algorithms are tailored for deep learning networks and do not possess strong portability.

\begin{table}[htbp]
  \centering
  \caption{Comparison with the existing quantization methods. Our LRQMM is an Data Free, operator-level quantization method and easy to combine with other methods}
    \begin{tabular}{cccc}
    \toprule
    Method & Data Free & Scope & \textcolor[rgb]{ .024,  .024,  .027}{Method Coupling} \\
    \midrule
    \midrule
    AdaDFQ\cite{ADASG} & Yes   & Network & Hard \\
    IntraQ\cite{Intraq} & No    & Network & Hard \\
    LQ-Nets\cite{zhang2018lq} & No    & Network & Hard \\
    AWQ\cite{MLSYS2024_42a452cb}   & Yes   & Operator & Hard \\
    Quantensor\cite{li2021unleashing} & Yes   & Operator & Easy \\
    LRQMM & Yes   & Operator & Easy \\
    \bottomrule
    \end{tabular}%
  \label{LRQMM_ADV}%
\end{table}%

To address these issue, we have developed LRQMM that keeps high computational accuracy in quantized matrix multiplication by combining residual compensation-based low-precision matrix multiplication with randomized SVD method, and the advantages of LRQMM are illustrated intuitively in Table.\ref{LRQMM_ADV}. Our contributions are as follows:

\begin{itemize}
    \item We propose LRQMM \textemdash  which introduces low-rank approximation based on RSVD into the quantization algorithm for residual compensation, proposing an improved low-precision quantized GEMM algorithm. LRQMM is a \textbf{data-free}, \textbf{operator level} quantization algorithm. It has \textbf{good portability} and is easy to be coupled with other quantization methods.

    \item We have presented the algorithm implementation process and conducted a time complexity analysis, demonstrating that LRQMM incurs additional cost at the \textbf{BLAS-2 level}. We analyze the theoretical error of the algorithm under the Frobenius norm, and our results show that LRQMM can bring several times the accuracy improvement.
    
    \item  We have implemented the algorithm on GPUs using state-of-the-art mathematical libraries, and have conducted a series of evaluations of LRQMM in GEMM and deep learning applications in comparison with several quantization methods, which proved that LRQMM has \textbf{1} to \textbf{3} orders of magnitude improvement of accuracy  under numerical calculation, and several times accuracy improvement in Deep Learning.
\end{itemize}

The rest of the paper is organized as follows: In Section.2, we introduce the RSVD algorithm and other related quantization algorithms. In Section.3, we propose the LRQMM algorithm and provide a detailed analysis of its time complexity and error bounds. In Section.4, we conduct a series of experiments to demonstrate the effectiveness of LRQMM. Finally, we provide a discussion and conclusion of LRQMM.

\section{Background and related works}

\subsection{Low rank approximate matrix multiplication}

\quad For any given \(m \times n\) matrix \( A \), there exists a decomposition that can be represented as \( A = U \Sigma V^T \) (SVD decomposition).
If only the largest \( k \) singular values and their corresponding left and right singular vectors are retained, it is possible to construct a \( k \)-rank matrix that closely approximates the original matrix \( A \), which shown in Eq.\ref{k-svd}, and this matrix \( A_k \) is the low-rank approximation of the original matrix \( A \) with minimal error:
\begin{equation}
\begin{split}
    A_k = U_k \Sigma_k V_k^T 
\end{split}
\label{k-svd}
\end{equation}
here, \( U_k \in \mathbb{R}^{m \times k} \) is an \( m \times k \) matrix composed of the first \( k \) columns of \( U \). \( \Sigma_k \) is a \( k \times k \) diagonal matrix containing the first \( k \) singular values. \( V_k^T \in \mathbb{R}^{n \times k} \) is a \( k \times n \) matrix composed of the first \( k \) rows of \( V \).

By employing the $k$-rank spanning matrix of SVD decomposition, it is straightforward to construct the required low-rank approximate matrices. However, the computational complexity required for a SVD decomposition is \( O(Cn^3) \) \footnote{$C$ is a constant that depends on the choice of SVD implementation algorithm, and due to the algorithm's inefficient parallelizability, it often necessitates a substantial amount of computational time.}.

To reduce computational time overhead, the randomized SVD (RSVD) approximation algorithm serves as an effective alternative. It has been widely applied in the field of computer vision \cite{ji2014gpu,zhou2014low,osawa2017accelerating} and machine learning \cite{kumar2016novel,guan2017matrix,mehta2017review}, demonstrating significant competitiveness in rapidly computing low-rank approximations of large matrices.

This algorithm is generally divided into the following two steps:

\begin{enumerate}
    \item Compute an approximate basis for the column space of \( A \in \mathbb{R}^{m \times n} \). Aiming to obtain a matrix \( Q \) with \( r \) orthogonal columns that approximates matrix \( A \). Formally, \( A_r \approx QQ^*A \), where \( Q^* \) denotes the conjugate transpose of \( Q \).
    \item Utilize the orthogonal matrix \( Q \) to calculate a much smaller rank-\( k \) matrix \( Q^*A \), and employ it to compute the desired matrix decomposition.
\end{enumerate}

In the case of RSVD, after obtain the matrix $Q$ through various sampling techniques\cite{frieze2004fast, musco2015randomized}. Then randomized SVD is the following Algorithm.\ref{algorithmofrsvd}:

\begin{algorithm}
  \SetAlgoLined
\KwData{$A,Q$(Intput matrix)}
  
  \KwResult{$U,\Sigma,V^{*}$(SVD decomposition matrix)}

    \tcc{Computes approximate matrix $B$}
    $\{B\}\leftarrow Q^{*}A$\;
    \tcc{Computes the SVD of $B$}
    $\{U'\Sigma V^{*}\}\leftarrow B$\;
    $\{U\} \leftarrow QU'$\;

    \Return $U,\Sigma,V^{*}$\;
    
  \caption{A simple process of Randomized SVD }
    \label{algorithmofrsvd}
\end{algorithm}

From the algorithm, it is evident that the computational requirements for RSVD are significantly reduced compared to the original SVD, with a time complexity of \( O(mnlog(r))+(m+n)r^2) \), and has the error satisfies:
\begin{equation}
\label{rsvderror}
\begin{split}
\mathbb E \Vert A-A_r\Vert \leq \left[ 1+4\sqrt{\frac{2min\{m,n\}}{r-1} }\right ]^{1/(2q+1)}\sigma_{r+1}.
\end{split}
\end{equation}
the proof of this theorem is complex; for specifics, one may refer to \cite{halko2011finding}.

Based on the RSVD approximation, for matrix multiplication \( C = A B \), where matrix \( A \) has dimensions \( m \times k \) and matrix \( B \) has dimensions \( k \times n \), we can approximate matrices \( A \) and \( B \) as Eq.\ref{AB-RSVD}:
\begin{equation}
\begin{split}
    A \approx A' = U_{r} \Sigma_{r} V_{r}^{T} \in \mathbb{R}^{m \times k}\\
    B \approx B' = W_{r} \Gamma_{r} Z_{r}^{T} \in \mathbb{R}^{k \times n}
\end{split}
\label{AB-RSVD}
\end{equation}
where \( A' \) and \( B' \) are the rank \( r \) approximations of \( A \) and \( B \), respectively. \( U_{r} \) and \( W_{r} \) are orthogonal matrices, \( \Sigma_{r} \) and \( \Gamma_{r} \) are diagonal matrices containing singular values, while \( V_{r} \) and \( Z_{r} \) are also orthogonal matrices spanned by right singular vectors.

To compute the product \( C' = A' \times B' \), we can leverage the properties of SVD. The product of the approximate matrices \( A' \) and \( B' \) is given by Eq.\ref{APPMM-C}:
\begin{equation}
\begin{split}
    C' = (U_{r} \Sigma_{r} V_{r}^{T}) (W_{r} \Gamma_{r} Z_{r}^{T}) = (U_{r}\Sigma_{r})( V_{r}^{T} W_{r} )(\Gamma_{r} Z_{r}^{T}) 
\end{split}
\label{APPMM-C}
\end{equation}

Thus, the original matrix multiplication is decomposed into these sub-matrix multiplication operations. Since the approximate rank \( r \) chosen is much smaller than \( k \), the resulting matrices after decomposition will be tall and thin. 

But this method is not universal, requiring the input matrix to exhibit low-rank characteristics, meaning that the largest singular values of the matrix should constitute the vast majority of the sum of all singular values.

\subsection{Low precision computation}
\quad To transform inputs of original precision for execution at low-integer precision, quantization operations are necessary. In the context of matrix multiplication \( C = AB \), let the matrices after low-precision quantization be denoted as \( A' \) and \( B' \).

According to the IEEE 754 standard\cite{kahan1996ieee}, a floating-point number is composed of three parts: the sign bit, the exponent bit, and the fraction (mantissa) bit. FP32, which is a single-precision floating-point, consists of 1 sign bit, 8 exponent bits, and 23 fraction bits. 

Quantizing to integer types is not straightforward due to the structure of floating-point numbers. Therefore, some mathematical transformations are required before converting floating-point numbers to integers.

A simple yet effective quantization method is described as follows: 
\begin{equation}
\begin{split}
    a_{int} = Q(a_{fp}, \lambda&) = \text{TypeCast}(\lambda a_{fp}, \text{intN}),  
    \\&\lambda = \frac{2^{N-1}-1}{a_{max}} . 
\end{split}
\label{quantA}
\end{equation}
here, \( \lambda \) is a scaling factor determined by the maximum absolute value \( a_{max} \) in the input data.

This scaling factor $\lambda$ is used to map the floating-point values to the representable range of integers. The process casts float values to the integer type \( \text{int}N \), where \( N \) is the number of bits used to represent the integer. This quantization approach helps to preserve the relative distribution of the original floating-point values within the limited range of the integer representation.

Through this approach, quantized matrix multiplication \( C = Quant(AB)\) can be represented as the three steps:

1. Quantization of Matrix \( A,B \):
\begin{equation}
\begin{split}
    (A|B)_{int} = Q(A|B, \lambda_{A|B}) = \text{round}(\lambda_{A|B} \cdot A|B)
\end{split}
\label{quantAB}
\end{equation}

2. Integer Matrix Multiplication:
\begin{equation}
\begin{split}
    C_{int} = A_{int} B_{int}
\end{split}
\label{INTGEMM}
\end{equation}

3. Dequantization of Result \( C_{int} \) to Floating-Point:
\begin{equation}
\begin{split}
    C_{Fp} = \widetilde{Q(C_{int}, \frac{1}{\lambda_A \lambda_B})} = \frac{1}{\lambda_A \lambda_B} \cdot C_{int}
\end{split}
\label{QUANTGEMM}
\end{equation}

This is a simple quantization method, which will introduce significant errors in practical applications. To reduce these errors, researchers have proposed various op-level data-free quantization techniques:

\begin{itemize}
    \item Vector-wise Quantization \cite{chen2021quantization} involves applying different scaling factors \( \lambda \) to each row of matrix \( A \) and each column of matrix \( B \). The goal is to reduce the variance of the input data by minimizing the difference between the maximum and minimum values, thereby improving the quantization accuracy. 
    \item Improved Bit-wise Quantization \cite{zhang2018lq,choi2021terngemm} leverages the bit-wise matrix operations supported by the latest GPUs. It allows for quantizing the inputs at different bit precisions, adjusting according to the input matrices to enhance quantization accuracy.
    \item Matrix Reordering \cite{han2020extremely} involves rearranging the input matrices, expanding matrix \( B \) along the rows, and transforming the matrix multiplication into a summation after the dot product. This enables the use of instruction-level algorithms that support low-precision quantization after the data has been reordered.
    \item Methods using clustering \cite{equitz1989new} and scaling \cite{dai2021vs,higham2019squeezing} to quantize floating-point values into integer values.
\end{itemize}

In addition to these various quantization methods, error compensation techniques can also be employed to enhance the result accuracy by performing supplementary calculations after the main computations are completed. Li\cite{li2021unleashing} introduces an iterative method in this process. Gu\cite{gu2024method} address the extensive redundant computations brought about by complete residual calculations, sparsity is introduced in this process.

Following the introduction in the section, the different quantization methods are primarily aimed at addressing errors under abnormal distributions, requiring adjustments based on the distribution of the input data, which results in poor generality.  On the other hand, residual error compensation methods can significantly improve accuracy,  but due to the additional computational load introduced by the residual terms, the complete residual compensation method is \textbf{N} times the computational load of the original quantization, which is often unacceptable in practice.  The sparsification method\cite{gu2024method} is limited by the speed of SPMM computations and tends to perform poorly on GPU platforms.

\section{METHODOLOGY}
\quad To achieve high computational accuracy while leveraging the high efficiency of low precision, we propose LRQMM (Low-rank Residual Quantized Matrix Multiplication). In the first section, we will introduce LRQMM and present its implementation process as in Fig.\ref{Eg of LRQMM}. In the second section, we will analyze the time complexity of LRQMM. In the third section, we will conduct an in-depth analysis of the algorithm's error, including an exploration of how errors are introduced and propagated through computation.

\subsection{Introduction to LRQMM Algorithm}
\quad In the quantization error method, let the error matrix be \( R_A \), and the matrix used for computation after low-precision quantization be \( A' \):
\begin{equation}
\begin{split}
    A^{Int} = Quant(A^{Fp});A^{Fp} = A^{'Fp} + R_A^{Fp}
\end{split}
\label{AFP}
\end{equation}

Applying the same process to matrix B, the matrix operation \( AB = C \) can be represented as Eq.\ref{AFPBF2}, where $(A|B)^{'Fp} = Dequant((A|B)^{Int})$:
\begin{equation}
\begin{split}
    A^{Fp} \cdot B^{Fp} = (A^{'Fp} + R_A^{Fp}) * (B^{'Fp} + R_B^{Fp})
\end{split}
\label{AFPBF2}
\end{equation}

The product of the two floating-point matrices can be expressed as the product of the quantized integer matrices multiplied by the scaling values \( \lambda \) of the two matrices. Therefore, adding the results of the four matrix multiplications yields the original matrix multiplication, which is the method of residual compensation.

\begin{equation}
\begin{split}
    A^{Fp}*B^{Fp} &=\underbrace{\frac{A^{Int}\cdot B^{Int}}{\lambda_{a}*\lambda_{b}}}_{T_1}+\underbrace{\frac{A^{Int}\cdot R_B^{Int}}{\lambda_{a}*\lambda_{Rb}}}_{T_2}  \\
    &\underbrace{+\frac{R_A^{Int}\cdot B^{Int}}{\lambda_{Ra}*\lambda_{b}} 
      +\frac{R_A^{Int}\cdot R_B^{Int}}{\lambda_{Ra}*\lambda_{Rb}}}_{T_2} = C^{Fp}
\end{split}
\label{gemm_r_split}
\end{equation}

According to Eq.\ref{gemm_r_split}, the complete residual compensation method can be considered as consisting of two parts. The first part involves a single quantized matrix multiplication of the original matrices. The second part encompasses three instances of residual compensation.
\begin{figure}
	\centering
\includegraphics[width=1\linewidth]{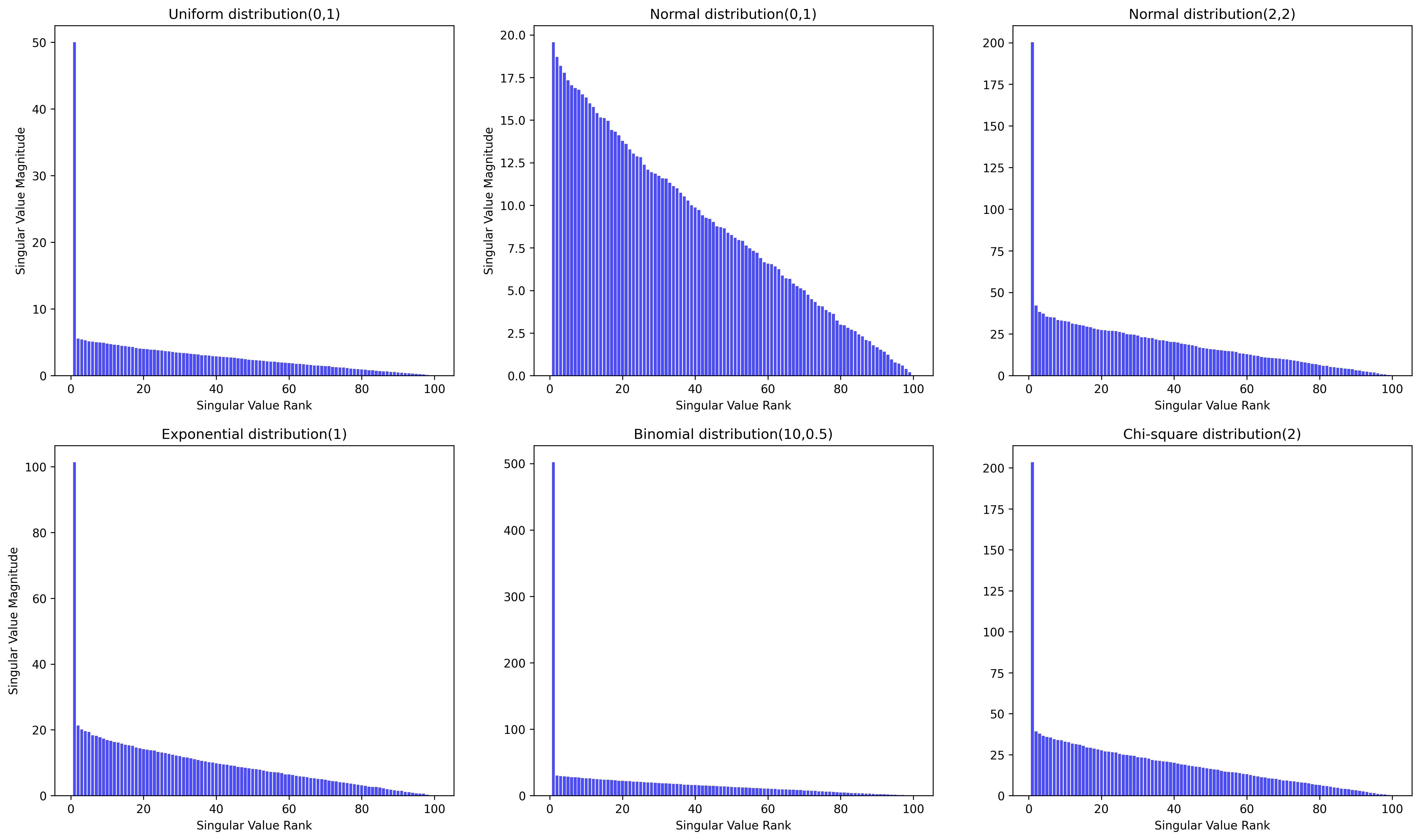}
	\caption{The distribution of singular values in matrices under different distributions, the dimensions are \( 100 \times 100 \).}
	\label{singular_value_bar_charts}
\end{figure}

For the second part, elements in $R_A, R_B$ are generated through the Eq.\ref{get_ra}: 
\begin{eqnarray}    \label{get_ra}
    ra^{Int}_{ij}= \lambda *a^{Fp}_{ij} - round(\lambda *a^{Fp}_{ij})
\end{eqnarray}

However, because each computation involves a complete dense matrix multiplication, a single complete residual calculation increases the computational load by three times.  The performance loss due to the final precision improvement after iteration is often not worth the trade-off. Thus, it occurs to us to reduce the computational load of this part through dimensionality reduction methods.

In the SVD algorithm, since only the matrix spanned by the top \( r \) singular values is selected, the overall error depends on the matrix spanned by the discarded singular values. In other words, the fewer singular values that are discarded, the smaller the overall error. The singular values, after being sorted, are illustrated in Fig.\ref{singular_value_bar_charts} for different matrices. It can be observed that matrices with a mean far from zero often have a large singular value and a set of small singular values, and such matrices tend to have significantly better precision in SVD decomposition.

For general quantization methods, the rounding operation is often employed, which results in the residual matrices \( R_A \) and \( R_B \) containing both positive and negative numbers with a mean close to zero. This does not guarantee the accuracy of low-rank matrix approximation multiplication. However, if a slight modification is made to the rounding operation, changing it from rounding to rounding down, that is
\begin{eqnarray}    \label{get_ra2}
    ra^{Int}_{ij}= \lambda *a^{Fp}_{ij} - \lfloor \lambda *a^{Fp}_{ij}\rfloor
\end{eqnarray}
which ensures that the residual matrices contain only positive numbers and are distributed randomly.

By employing this approach,  we can utilize low-rank approximate multiplication in the three residual compensation parts instead of the original full-size matrix multiplication, thereby accelerating the computation.

In the case of the three instances of low-rank approximate matrix multiplication, due to the properties satisfied by \( R_A \) and \( R_B \), it is possible to approximate the entire residual matrices with a very small rank. Consequently, the approximation terms involving \( R_A \) and \( R_B \) in the matrix multiplications become tall and skinny matrix multiplication, which can be approximated as matrix-vector multiplication, and the three instances can be computed using either the original precision or half precision. We show the algorithm flow of LRQMM in Fig.\ref{Eg of LRQMM}.

\subsection{Time complexity analysis}
\quad We present the algorithm pseudocode of LRQMM in Algorithm.\ref{algorithmRLRRAMM}, assume that the input and output matrices are both of size $N*N$. The algorithm flow is shown as follows:

\begin{figure*}
	\centering
\includegraphics[width=1\linewidth]{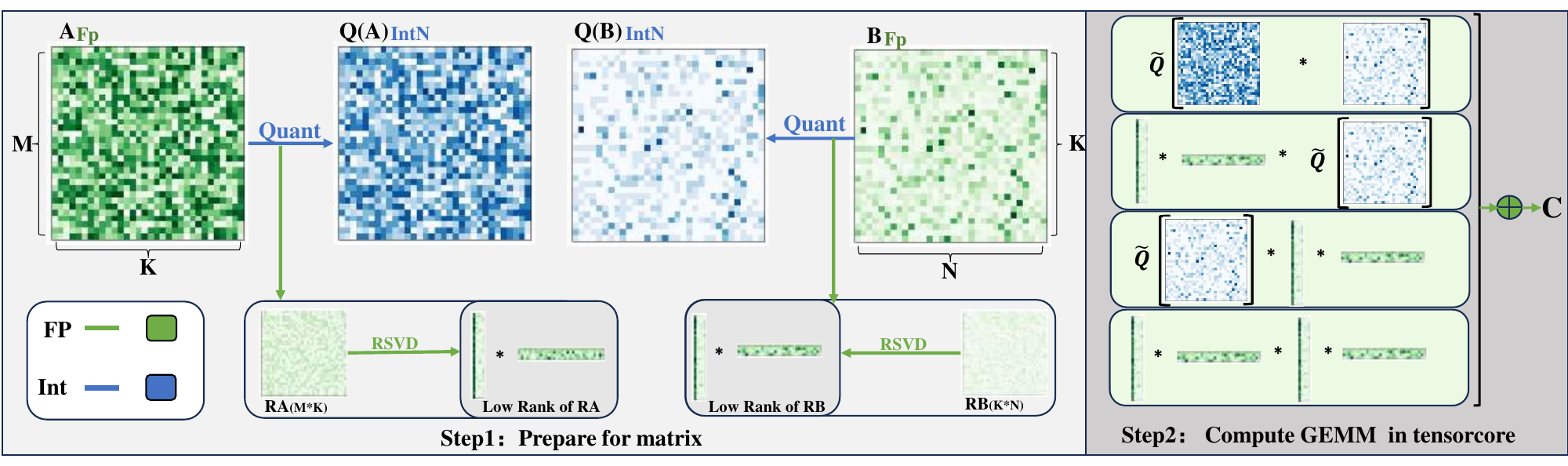}
	\caption{Illustration of the LRQMM process. We have extracted a $32x32$ output from \textit{RESNET} convolutional layer as the data source for visualization. $Q$ represents the quantization operation, and \( \widetilde{Q} \) represents the dequantization operation.
}
	\label{Eg of LRQMM}
\end{figure*}
\begin{enumerate}
    \item Normal low-precision quantized matrix multiplication: the low precision matrix $C_{int}$ under N integer is calculated by direct quantization operation, and the original precision result $C_F$ is obtained by dequantization  operation (lines 1-3).
    \begin{itemize}
        \item Firstly, $A$ and $B$ matrices are quantized, and the quantized matrix multiplication of $A_{int}$,$B_{int}$, which is of complexity $O(2*N^2)$.
        \item And then through $A_{int}$, $B_{int}$ matrix multiplication to get the quantized result matrix, which is of complexity $O(N^3)$.
        \item Dequantization of the quantization result matrix $C_{int}$ is of complexity $O(N^2)$.
    \end{itemize}
    \item Low-rank decomposition: In this step, the residual matrix $R_A, R_B$ of $A$ and $B$ are calculated by dequantization operation (lines 4-5). Subsequently, the obtained residual matrices are subjected to low-rank decomposition in RSVD, yielding six sub-matrices as a result of the decomposition (lines 6-7).
    \begin{itemize}
        \item The residual matrix $R_A$, $R_B$ is obtained by dequantization of the quantization matrix  $A_{int}$,$B_{int}$, which is of complexity $O(2*N^2)$.
        \item Applying low-rank decomposition to \( R_A \) and \( R_B \) ,which is of complexity $O(2N^2log(r)+4Nr^2)$.
    \end{itemize}    
    \item Computing a series of tall and skinny low-rank matrices multiplication of the residual compensation terms \( RC_F^1 \), \( RC_F^2 \) and \( RC_F^3 \) (lines 8-12).
    \begin{itemize}
        \item Computing the product of diagonal matrices to get ${\widetilde{U_r},\widetilde{Z_r'}}$, which is of complexity $O(2N^2)$.
        \item Computing $RC_F^1,RC_F^2,RC_F^3$, which is of complexity $O(6(rN^2))$.
        \item Adding three residual terms to the original quantized matrices for error compensation, which is of complexity $O(3*N^2)$.
    \end{itemize}      
\end{enumerate}
\begin{algorithm}
  \SetAlgoLined
\KwData{$A,B,C$(Input matrix);$\alpha$,$\beta$(Scalar);\newline N(Quant bit),$T_F$(Origin precision ),$r$(approximate rank)}
  
  \KwResult{matrix D}

    \tcc{Computes quantized GEMM}
    $\{A_{int},B_{int}\}\leftarrow Quant(\{A,B\},N)$\;
    $C_{int} =A_{int} B_{int}$\;
    $C_{F} \leftarrow Dequant(C_{int},T_F)$\;

    \tcc{Calculate the residual matrix}
    $\{A_{F},B_{F}\} \leftarrow Dequant(\{A_{int},B_{int}\},T_F)$\;
    $\{R_A,R_B\} \leftarrow \{A,B\} - \{A_{F},B_{F}\}$\;

    \tcc{The residual matrix decomposed in low rank}
    $\{U_{r},\Sigma_{r},V_{r}^{T}) \}\leftarrow RSVD(R_A,r)$\;
    $\{W_{r},\Gamma_{r},Z_{r}^{T}) \}\leftarrow RSVD(R_B,r)$\;

    \tcc{Calculate the product of the diagonal matrix $\Sigma_{r}, \Gamma_{r}.$}
    ${\widetilde{U_r},\widetilde{Z_r'}}\leftarrow ( \{U_{r} \cdot \Sigma_{r},\Gamma_{r}\cdot Z_{r}^{T}\} )$\;

    \tcc{Calculate three residual terms}
    $RC_F^1 = \widetilde{U_r} \cdot (V_{r}^{T} \cdot B_{F}) $\;
    $RC_F^2 = (A_{F} \cdot W_{r}) \cdot \widetilde{Z_r'} $\;
    $RC_F^3 = \big(\widetilde{U_r} \cdot( V_{r}^{T}  \cdot W_{r}) \big) \cdot \widetilde{Z_r'}$\;

    \tcc{Error compensation}
    $C_{F} = C_{F} + RC^{1}_{F} + RC^{2}_{F} + RC^{3}_{F}$\;

    \Return $D = \alpha*C_{F} + \beta D$\;
    
  \caption{Algorithms of compute GEMM $D=\alpha A\cdot B+\beta D$ in LRQMM }
  \label{algorithmRLRRAMM}
\end{algorithm}

To summarize, LRQMM has a time complexity of \( O(10+6r+2log(r)N^2 + N^3) \). Compared to the direct quantization method, which is of complexity $O(3*N^2+N^3)$. The additional complexities introduced are all of \( O(CN^2) \) when $r$ is small enough. Overall, the extra computational overhead is within an acceptable range.

\subsection{Error analysis of LRQMM}

\quad In this section,  we will present the error analysis of LRQMM through a series of derivations and compare it with the error of direct quantization.  Our goal is to derive an error expression that consists solely of the matrix dimensions, quantization parameter, mean, and variance without singular value, which has more practical reference significance.

Before commencing the proof, we first present several foundational theorems.

\subsubsection{Preliminaries of norm and eigenvalue}\

\textbf{Minkowski inequality:}
\quad \textit{If $A \in \mathbb R^{m*n}, B\in \mathbb R^{m*n}$:}
\begin{equation}
\label{Minkowski}
\begin{split}
||A+B||\leq||A||+||B||
\end{split}
\end{equation}

\textbf{Cauchy-Schwarz inequality:} \textit{If $A \in \mathbb R^{m*k}, B\in \mathbb R^{k*n}$:}
\begin{equation}
\label{Cauchy}
\begin{split}
||AB||\leq \Vert A\Vert \Vert B \Vert
\end{split}
\end{equation}

\textbf{Frobenius norm inequality:} \textit{If $A \in \mathbb R^{m*n}$ and $ (m \ge n)$, has SVD decomposition $A = \sum_{i=1}^n{\sigma_{i}u_iv_i^T}$, then}
\begin{equation}
\label{Frobenius_inequality}
\begin{split}
 \Vert A\Vert_F  \leq \sqrt{n}\sigma_1
\end{split}
\end{equation}

\textbf{Marchenko-Pastur Law\cite{baik2005phase}:} 
Suppose that $\{x_{jk}, j,k = 1,2,\dots\}$ is a double array of \textit{iid} (independent identically distributed) complex random variables with mean zero and variance $\sigma_2$. Write $x_j = (x_{1j}, \dots ,x_{pj})'$ and $X = (x_1,\dots,x_n)$. The sample covariance matrix is defined by
\begin{equation}
\label{Marchenko-Pastur Law}
\begin{split}
S = \frac{1}{n-1}\sum_{k=1}^n(x_k-\overline{x})(x_k-\overline{x})^*
\end{split}
\end{equation}
where $\overline{x} = \frac{1}{n}\sum_{j=1}^n{x_j}$. However, in most cases of spectral analysis of large dimensional random matrices, the sample covariance matrix is simply defined as
\begin{equation}
\label{SampleVar_2}
\begin{split}
S = \frac{1}{n}\sum_{k=1}^nx_kx_k^* = \frac{1}{n}XX^*
\end{split}
\end{equation}

Assume that $p/n \stackrel{n\rightarrow \infty}{\longrightarrow} y \in (0,1]$. Then we have the $p_y(x)$ deterministic to measure whose density of eigenvalue is given by
\begin{equation}
\label{mplawpx}
\begin{split}
p_y(x) =     
\begin{cases}
\frac{1}{2\pi xy\sigma^2}\sqrt{(b-x)(x-a)} &,if a\leq x\leq b,\\
0 &,otherwise
\end{cases}
\end{split}
\end{equation}
where $a = \sigma^2(1-\sqrt{y})^2$ and $b = \sigma^2(1+\sqrt{y})^2$.

Subsequently, we derive several corollaries based on these theorems.
\subsubsection{Corollaries of F-norm in SVD and quantization GEMM}\

\textbf{The error of low-rank approximation using SVD:}  
\textit{If $A \in \mathbb R^{m*n}$ with rank $p$ has SVD decomposition, take the first \( r(r<p) \) terms to approximate matrix \( A \). Then, the factorization satisfies:}

\begin{equation}
\label{svdfinq}
\begin{split}
\mathbb E||A - A_r||_F \leq \sigma_{k+1}\sqrt{p-r} 
\end{split}
\end{equation}
\textit{Proof.} Due to 
\begin{equation}
\label{EAARF}
\begin{split}
||A - A_r||_F &= ||\sum_{i=1}^{p} \sigma_i u_i v_i^T -  \sum_{i=1}^{r} \sigma_i u_i v_i^T||_F \\    
            &= ||\sum_{i=r+1}^{p} \sigma_i u_i v_i^T||_F = \sqrt{\sigma_{r+1}^2 + \dots + \sigma_{p+1}^2}\\
            &\leq  \sigma_{r+1}\sqrt{p-r}
\end{split}
\end{equation}

\textbf{Quantization error of uniformly dithered scalar quantizer:}
\textit{For a scalar $x \in \left [-M, +M \right ]$, denote the quantization error of direct scalar quantizer with a bit-budget of $d$ bits as $\epsilon = Quant(x) - x$, denote $\frac{2^{d-1}-1}{M}$ as $\lambda$. The mean and variance of $\epsilon$ satisfies}
\begin{equation}
\label{direct_quant_E}
\begin{split}
\mathbb E(\epsilon) = 0\ and\ Var(\epsilon) \leq \lambda^{-2}
\end{split}
\end{equation}

\textbf{Quantization error analysis of matrix:} \textit{If $A \in \mathbb R^{m*n}$, and we have quantized martrix $\widetilde{A} = Dequant(Quant(A))$. Then we have}
\begin{equation}
\label{quantEA}
\begin{split}
\Vert A-\widetilde{A} \Vert_F \leq \sqrt{mn}\lambda^{-1} \ and \ \mathbb E(\Vert A-\widetilde{A}\Vert_F) = 0
\end{split}
\end{equation}
\textit{Proof.} 
\textbf{(1).} For the first term, due to the quantization operation, the error introduced for each element \( a_{ij} \) in matrix \( A \) can be represented as:
\begin{eqnarray}    \label{countAint}
    ra^{Int}_{ij}= \lfloor\lambda *a^{Fp}_{ij}\rfloor - \lambda *a^{Fp}_{ij} 
\end{eqnarray}
where $\lfloor.\rfloor$is the integer type casting. Since $ra^{Int}_{ij}\leq1$(Due to the rounding operation, the error in integer will not exceed 1).
Thus 
\begin{equation}
\label{RAFPIJ}
\begin{split}
ra^{Fp}_{ij}=TypeCast((ra^{Int}_{ij}*\lambda^{-1},Float)\leq \lambda^{-1}
\end{split}
\end{equation}
\quad The errors \( ra_{ij} \) for each element in matrix \( A \) can be concatenated to form an matrix \( R_A \in \mathbb R^{m*n}\). Then we have 
\begin{equation}
\label{AWAF}
\begin{split}
\Vert A-\widetilde{A} \Vert_F &= \Vert R_A \Vert_F= \sqrt{\sum_{i=1}^m\sum_{j=1}^n {ra_{ij}^{Fp}}^2}
        \leq \sqrt{mn}\lambda^{-1}
\end{split}
\end{equation}
\quad \textbf{(2).}For the second term, due to the quantization operation is performed on the entire matrix, we can expand the matrix $A$ into vector. And then we can prove it by applying the previous Eq.\ref{direct_quant_E}.

\textbf{Error analysis of Quantized matrix multiplication:} 
\textit{If $C=A*B, \in \mathbb R^{m*n}$ where $A \in \mathbb R^{m*k}, B\in \mathbb R^{k*n}$. Assuming that matrices \( A \) and \( B \) follow the same distribution and $k \leq m\leq n$. Denote the singular value of matrix $A,B$ is $\sigma,\gamma$. For quantized matrix multiplication, $\widetilde{C}=\widetilde{A}*\widetilde{B}$ with $\lambda_1,\lambda_2$. We have}
\begin{equation}
\label{er_quant}
\begin{split}
\mathbb E(\Vert C-\widetilde{C} \Vert_F) &\leq k(\sigma_1\lambda_2^{-1}\sqrt{n}+\gamma_1\lambda_1^{-1}\sqrt{m} +\lambda_1^{-1}\lambda_2^{-1}\sqrt{mn})
\end{split}
\end{equation}
\textit{Proof.}
\begin{equation}
\label{ECCF}
\begin{split}
\mathbb E(\Vert C-\widetilde{C} \Vert_F)  &= E(\Vert AB-\widetilde{A}\widetilde{B} \Vert_F) \\
        &= E(\Vert (\widetilde{A}+R_A)(\widetilde{B}+R_B)-\widetilde{A}\widetilde{B} \Vert_F) \\
        &= E(\Vert R_A\widetilde{B}+\widetilde{A}R_B+ R_A R_B \Vert_F) 
\end{split}
\end{equation}
Subsequently, by the Minkowski and Cauchy-Schwarz inequality presented in Eq.\ref{Minkowski},\ref{Cauchy}
\begin{equation}
\label{EMINKOWSKI_bef}
\begin{split}
\mathbb  E(\Vert C-\widetilde{C} \Vert_F) \leq \mathbb E(\Vert R_A\Vert_F \Vert\widetilde{B}\Vert_F+\Vert \widetilde{A}\Vert_F \Vert R_B \Vert_F + \Vert R_A \Vert_F \Vert R_B \Vert_F)
\end{split}
\end{equation}

Due to Eq.\ref{quantEA}. The $\Vert R_A \Vert_F= \Vert A-\widetilde{A} \Vert_F \leq \sqrt{mk}\lambda_1^{-1}$, the same way $\Vert R_B \Vert_F \leq \sqrt{kn}\lambda_2^{-1}$. And due to the quantization operation, the maximum absolute value of the quantized matrix will not exceed the maximum absolute value of the original matrix, and from Eq.\ref{Frobenius_inequality}, we have:
\begin{equation}
\label{AFBF}
\begin{split}
\Vert \widetilde{A} \Vert_F \leq \Vert A \Vert_F \leq \sigma_1 \sqrt{k}, \Vert \widetilde{B} \Vert_F \leq \Vert B \Vert_F \leq \gamma_1\sqrt{k}\\
\end{split}
\end{equation}
incorporating these components into Eq.\ref{EMINKOWSKI_bef}, then we can finish the proof.
\begin{equation}
\label{ECCF2}
\begin{split}
\mathbb E(\Vert C-\widetilde{C} \Vert_F)  &\leq k(\sigma_1\lambda_2^{-1}\sqrt{n}+\gamma_1\lambda_1^{-1}\sqrt{m}+\lambda_1^{-1}\lambda_2^{-1}\sqrt{mn})
\end{split}
\end{equation}

\textbf{Better inequation of Frobenius norm:} 
\textit{If $iid$ matrix $A \in \mathbb R^{m*n}, m>n$ with mean $\mu$ and variance $s^2$. And let $B = A-\mu$. Denote $\kappa_1 =|s(1+\sqrt{n/m})|, \kappa_2 = |\mu|\sqrt{mn}$. The largest singular value $\sigma_1$ of $A$ can be estimated from:}
\begin{equation}
\label{maxlambda_copy}
\begin{split}
\sigma_1 \approx  max( \kappa_1, \kappa_2), \sigma_2 \approx min(\kappa_1, \kappa_2)
\end{split}
\end{equation}
And a more precise F-norm can be expressed as:
\begin{equation}
\label{percise_fnorm}
\begin{cases}
\Vert A \Vert_F \leq \sqrt{n}\sigma_1, \ \ \ \ \ \ \ \ \ \ \ \ (\kappa_1 \geq \kappa_2) \dots T_1\\
\Vert A \Vert_F \leq \sigma_1 + \sqrt{n}\sigma_2,\ \ \ \ \ (\kappa_1 < \kappa_2)  \dots T_2 \\
 \end{cases}
\end{equation}
\textit{Proof.} 
Since the singular values of a matrix are composed of the square roots of the eigenvalues of the product of the matrix itself and its transpose. Given that \( B \) satisfy the $iid$ condition, with a mean of 0 and variance $\sigma_2$. Therefore, the eigenvalues of matrix \( B \) naturally conform to the distribution function in Eq.\ref{mplawpx}. Then, the eigenvalues of the sample covariance matrix of \( B \) are at most \( b = \sigma^2(1 + \sqrt{y})^2 \). 

Assuming matrix \( C \) is a rank-one matrix spanned by the scalar \( \mu \), the singular value of matrix \( C \) are also the square roots of the eigenvalue of \( C^T C \) which is $\sqrt{\sum_{i=1}^n\sum_{i=1}^mc_{ij}^2} = |\mu|\sqrt{mn}$.
Thus, the largest singular value of matrix \( A \) can be approximated as the maximum singular value among those of matrices \( B \) and \( C \), and $\sigma_2$ is the smaller one. That is, Eq.\ref{maxlambda_copy}. 

When \( \kappa_1 \geq \kappa_2 \) the singular values are continuous, the Frobenius norm of the matrix can be directly expressed by \( \sigma_1 \). But when \( \kappa_1 < \kappa_2 \), since the first singular value exhibits a discontinuity compared to the others, the estimation of the matrix's Frobenius norm cannot be directly based on \( \sigma_1 \), then $\Vert A \Vert_F = \sqrt{\sigma_{1}^2 + \dots + \sigma_{n}^2} \leq \sigma_1 + \sqrt{\sigma_2^2 + \dots + \sigma_{n}^2} \leq \sigma_1 + \sqrt{n}\sigma_2.\\$

\

Now, we have presented all the necessary prerequisite corollaries. To facilitate error analysis, we make the following assumptions: $A\in \mathbb R^{m*k}, B\in \mathbb R^{k*n}, k \leq m \leq n $ and both satisfy same $iid$ condition with mean $\mu$ and variance $\sigma$ and maximum absolute value $max$. The low-rank approximate rank $r < k-1$.

\subsubsection{Further error analysis of Quant GEMM}
Apply Eq.\ref{maxlambda_copy} to the error of Quant GEMM in Eq.\ref{er_quant} . Now there are two conditions:
\begin{enumerate}
    \item When \( \kappa_1 \geq \kappa_2 \). Apply $T_1$ in Eq.\ref{percise_fnorm}. The matrix mean does not significantly affect the singular values, which still approximately follow the Marchenko-Pastur Law. In this case
    \begin{equation}
    \label{Er_case1_p}
    \begin{split}
    \Vert C-\widetilde{C} \Vert_F &\leq k\lambda^{-1}(\sqrt{n}s(1+\sqrt{k/m}) \\
    &+\sqrt{m}s(1+\sqrt{n/k})+\sqrt{mn}\lambda^{-1})
    \end{split}
    \end{equation}

    \item  When \( \kappa_1 < \kappa_2 \). Apply $T_2$ in Eq.\ref{percise_fnorm}. Then the error of direct quant GEMM approximation constrained by mean and variance can be expressed as Eq.\ref{Er_case2_p}:
    \begin{equation}
    \label{Er_case2_p}
    \begin{split}
    \Vert C-\widetilde{C} \Vert_F &\leq k\lambda^{-1}(s(\sqrt{n}+\sqrt{m}+\sqrt{\frac{nk}{m}}+ \sqrt{\frac{mn}{k}})\\
    &+2|\mu|\sqrt{mn})+\lambda^{-2}\sqrt{mn}
    \end{split}
    \end{equation}

\end{enumerate} 

\subsubsection{Error analysis of LRQMM}: If $A,B $, and has quantized martrix $\widetilde{A}|\widetilde{B} = {Dequant(Quant(A|B))}$ with quant parameter $\lambda$. Let the quantized residual matrix be denoted as \( R_A|R_B = A|B-\widetilde{A}|\widetilde{B} \) and \( R_{R_A}|R_{R_B} = R_A|R_B-R_{Ar}|R_{Br} \). From Eq.\ref{direct_quant_E}, we have:
\begin{equation}
\label{quantE}
\begin{split}
\mu = \mathbb E(R_A|R_B) \leq \lambda^{-1}\ ,  s^2 = Var(R_A|R_B) \leq\lambda^{-2}
\end{split}
\end{equation}
while the quantization operation uses rounding down, for the residual matrices \( R_A \) and \( R_B \), it holds that \( \kappa_1^{R_A} << \kappa_2^{R_A} \) when the matrix scales are large enough. Consequently: $\sigma_1^{R_A }>> \sigma_2^{R_A} > \sigma_{r+1}^{R_A}$, which is same to $R_B$.

Denote $C_L = LRQMM(A,B)$, then:
\begin{equation}
\label{quantCE}
\begin{split}
\Vert C-C_L \Vert_F &\leq \mathbb E(\Vert R_{R_A}\Vert_F \Vert\widetilde{B}\Vert_F+\Vert \widetilde{A}\Vert_F \Vert R_{R_B} \Vert_F \\
    &+ \Vert R_{A} \Vert_F \Vert R_{R_B} \Vert_F) 
\end{split}
\end{equation}

Apply Eq.\ref{maxlambda_copy}:
\begin{enumerate}
    \item When \( \kappa_1 \geq \kappa_2 \). Apply $T_1$ in Eq.\ref{percise_fnorm}.The matrix mean does not significantly affect the singular values, which still approximately follow the Marchenko-Pastur Law distribution. Then the error of LRQMM is:
    \begin{equation}
    \label{Er_case1_lrqmm}
    \begin{split}
    \Vert C-C_L \Vert_F &\leq \lambda^{-1}\sqrt{k(k-r)}((2s+\lambda^{-1})(1+\sqrt{\frac{k}{m}})
    \\&(1+\sqrt{\frac{n}{k}})+\lambda^{-1}\sqrt{m}(1+\sqrt{\frac{n}{k}})) \dots P_1
    \end{split}
    \end{equation}

    \item  When \( \kappa_1 < \kappa_2 \). Apply $T_2$ in Eq.\ref{percise_fnorm}. Then the error of LRQMM can be expressed as Eq.\ref{Er_case2_lrqmm}:
    \begin{equation}
    \label{Er_case2_lrqmm}
    \begin{split}
    \Vert C-C_L \Vert_F &\leq P_1 + \lambda^{-1}|\mu|\sqrt{k(k-r)}\\ 
    &(\sqrt{n}+\sqrt{m}+\sqrt{\frac{nk}{m}}+ \sqrt{\frac{mn}{k}})
    \end{split}
    \end{equation}

\end{enumerate} 

\subsubsection{Comparative error analysis} Now, we will conduct a more detailed comparative analysis to determine the extent of the precision enhancement of the LRQMM algorithm. To facilitate the simplification of formulas and analysis of results, we assume that matrices A and B are both square matrices, with \( m = n = k \). Still we have two conditions:
\begin{enumerate}
    \item when \( \kappa_1 \geq \kappa_2 \):
    \begin{equation}
    \label{Er_case1_RELA1}
    \begin{split}
    \frac{\Vert C-C_L \Vert_F}{\Vert C- \widetilde{C} \Vert_F} &= \frac{\sqrt{k(k-r)}(8s\lambda^{-1}+\lambda^{-2}(2\sqrt{k}+4))}  {4sk\lambda^{-1}\sqrt{k}+k\lambda^{-1}} \\
    & <  \frac{\lambda^{-1}\sqrt{k-r}}{2s\sqrt{k}} + \frac{2\sqrt{k-r}}{k} + \frac{\lambda^{-1}\sqrt{k-r}}{sk}  \\
    \end{split}
    \end{equation}    
    Notice that $\lambda = \frac{2^{d-1}-1}{max}$ and $s<max$, then $\lambda s < 2^{d-1}-1$. When $k$ is large, the magnitude of the last two terms is close to zero, while the first term is approximately a constant $\frac{1}{2^d - 2} < 1 $.

    \item when \( \kappa_1 < \kappa_2 \):
    \begin{equation}
    \label{Er_case2_RELA1}
    \begin{split}
    \frac{\Vert C-C_L \Vert_F}{\Vert C-\widetilde{C} \Vert_F} &= \frac{\lambda^{-1}\sqrt{k(k-r)}(8s+4\sqrt{k}|\mu| + \lambda^{-1}(2\sqrt{k}+4))} {k\lambda^{-1}(2\sqrt{k}(2s+|\mu|\sqrt{k})+\lambda^{-2})} \\
    &< \frac{\sqrt{k-r}(2|\mu|+\lambda^{-1})}{k|\mu|} + \frac{\sqrt{k(k-r)}(8s+4\lambda^{-1})}{2k^2|\mu|}
    \end{split}
    \end{equation}   
    In this inequality, as \( k \) becomes sufficiently large, the right-hand side approaches \( k^{-\frac{1}{2}} \).

\end{enumerate}
\quad Now, we have derived the error rate forms of LRQMM relative to direct quantization under two cases. In the first case, the error ratio approaches a constant, which is related to the number of quantization bits; that is, the higher the number of quantization bits, the higher the relative accuracy of LRQMM. This is consistent with the experimental results in Table.\ref{LRQMMM_int4},\ref{LRQMMM_int8}. In the second case, the error ratio approaches \( k^{-1/2} \), which means that LRQMM has a significant precision improvement, and this is also consistent with the curves in Fig.\ref{rank_precision} and the results in Table.\ref{LRQMMM_int4},\ref{LRQMMM_int8}.

\section{Evaluation}
\quad To evaluate the effectiveness of LRQMM, we conducted experimental tests from multiple aspects, including precision testing under different scales and distributions, as well as performance testing of the algorithm.

 In this section, we evaluate the performance of LRQMM on \textit{Nvidia-A100 GPU} platforms. Low-precision and single-precision matrix multiplications from \textit{cutlass3.5.0} and \textit{cuBLAS} were utilized, along with the RSVD function from \textit{cuSolver}.

\subsection{Precision test}
\quad The following accuracy tests for error are all based on the relative error in the Frobenius norm.

\begin{figure}
	\centering
\includegraphics[width=1\linewidth]{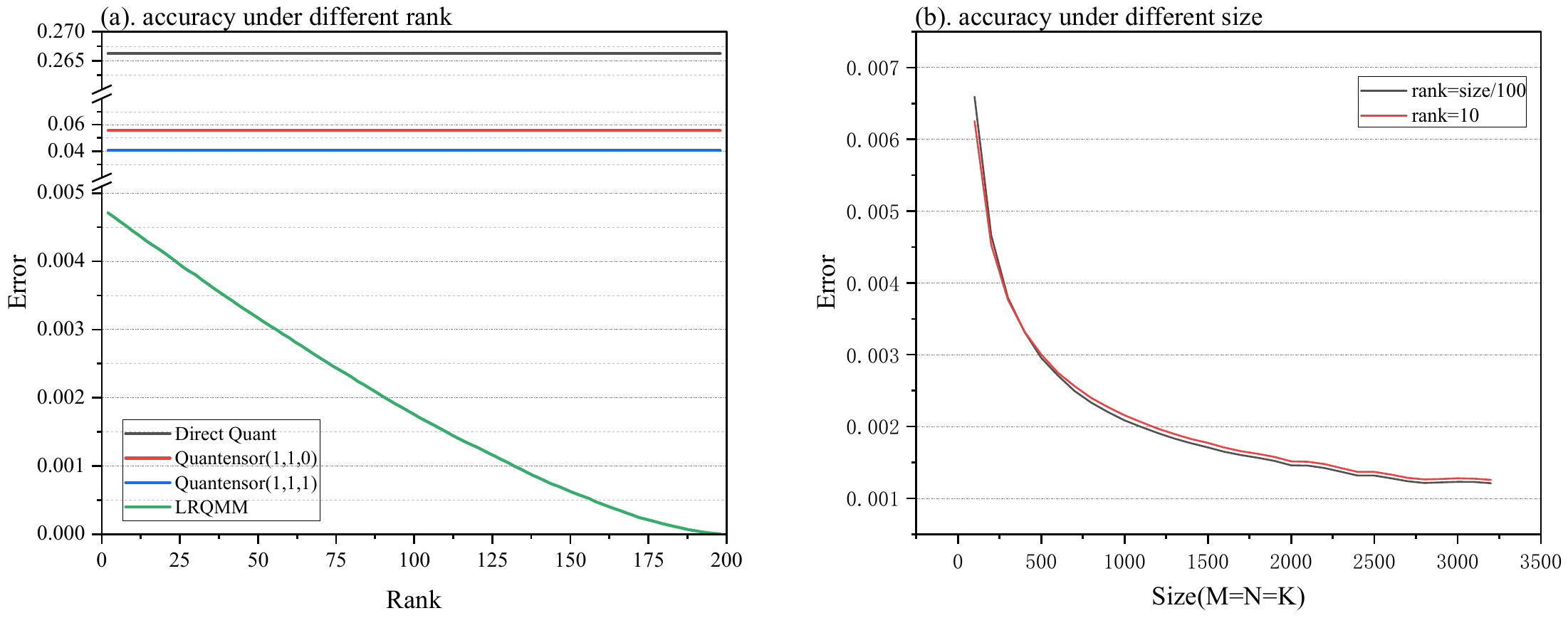}

	\caption{The algorithm's accuracy under different ranks of approximation (a), where the matrix used for testing is the uniform distribution matrix of size $200^3$.  Accuracy under different matrix scales (b). 
 }
	\label{rank_precision}
\end{figure}

\textbf{Different approximate rank:}
The algorithm, under the condition of identical matrix dimensions, utilizes various rank approximations, with accuracy depicted in Fig.\ref{rank_precision}, (a). From the result, the following two conclusions can be drawn:
\begin{itemize}
    \item The error of the low-rank approximation algorithm with residual compensation decreases as the approximate rank increases.
    \item The relationship between the increase in rank and the error is nearly linear; the degree of error reduction is within an order of magnitude as the rank increases from 1 to maximum value. In other words, using a small rank can yield relatively accurate results
\end{itemize}

\textbf{Different dimension:}
In this experiment, we fix the matrix approximation rank \( r = 10 \) and \( r = size/100 \) to test the variation of the algorithm's error as the matrix scale increases, as shown in Fig.\ref{rank_precision}, (b). It can be observed from the figure that the accuracy of the algorithm does not deteriorate with the enlargement of the matrix scale; on the contrary, the accuracy improves with the increase in matrix size. Through this experiment and the previous one, we can see that the algorithm can achieve good results at a very low approximation rank, and it can still maintain excellent numerical stability as the matrix scale increases.

\begin{table}[htbp]
  \centering
  \caption{The relative error of the algorithm under Int4 quantization under F-Norm. Where DQ is short for Direct Quant. QT is short for Quantensor.}
    \begin{tabular}{rrrrr}
    \toprule
    Type  & DQ & QT-110    & QT-111 & LRQMM \\
    \midrule
    Normal(0,1)     & 5.69E-1 & 1.98E-1 & 1.01E-1 & 2.10E-1 \\
    Uniform(0,1)    & 2.59E-1 & 5.48E-2 & 3.94E-2 & 1.46E-3 \\
    Uniform(-1,1)    & 2.39E-1 & 4.76E-2 & 3.58E-2 & 1.00E-1 \\
    Exponent(4)     & 9.11E-1 & 6.00E-1 & 2.68E-1 & 9.91E-3 \\
    ChiSquare(1)     & 9.52E-1 & 7.18E-1 & 4.15E-1 & 4.72E-2 \\
    Poisson(10)     & 3.68E-1 & 9.10E-2 & 5.97E-2 & 9.55E-4 \\
    \bottomrule
    \end{tabular}%
  \label{LRQMMM_int4}%
\end{table}%
\begin{table}[htbp]
  \centering
  \caption{The relative error of the algorithm under Int8 quantization F-Norm. Where DQ is short for Direct Quant. And QT is short for QuantTensor.}
    \begin{tabular}{rrrrr}
    \toprule
    Type  & DQ & QT-110    & QT-111 & LRQMM \\
    \midrule
    Normal(0,1)     & 4.05E-2 & 7.98E-4 & 3.24E-4 & 1.15E-2 \\
    Uniform(0,1)    & 1.56E-2 & 1.80E-4 & 1.33E-4 & 8.14E-5 \\
    Uniform(-1,1)    & 1.38E-2 & 1.56E-4 & 1.09E-4 & 5.52E-3 \\
    Exponent(4)     & 1.11E-1 & 4.26E-3 & 9.35E-4 & 5.86E-4 \\
    ChiSquare(1)     & 2.17E-1 & 1.56E-2 & 2.63E-3 & 3.48E-3 \\
    Poisson(10)     & 2.22E-2 & 3.18E-4 & 1.93E-4 & 4.89E-5 \\
    \bottomrule
    \end{tabular}%
  \label{LRQMMM_int8}%
\end{table}%

\begin{figure*}
	\centering
\includegraphics[width=1\linewidth]{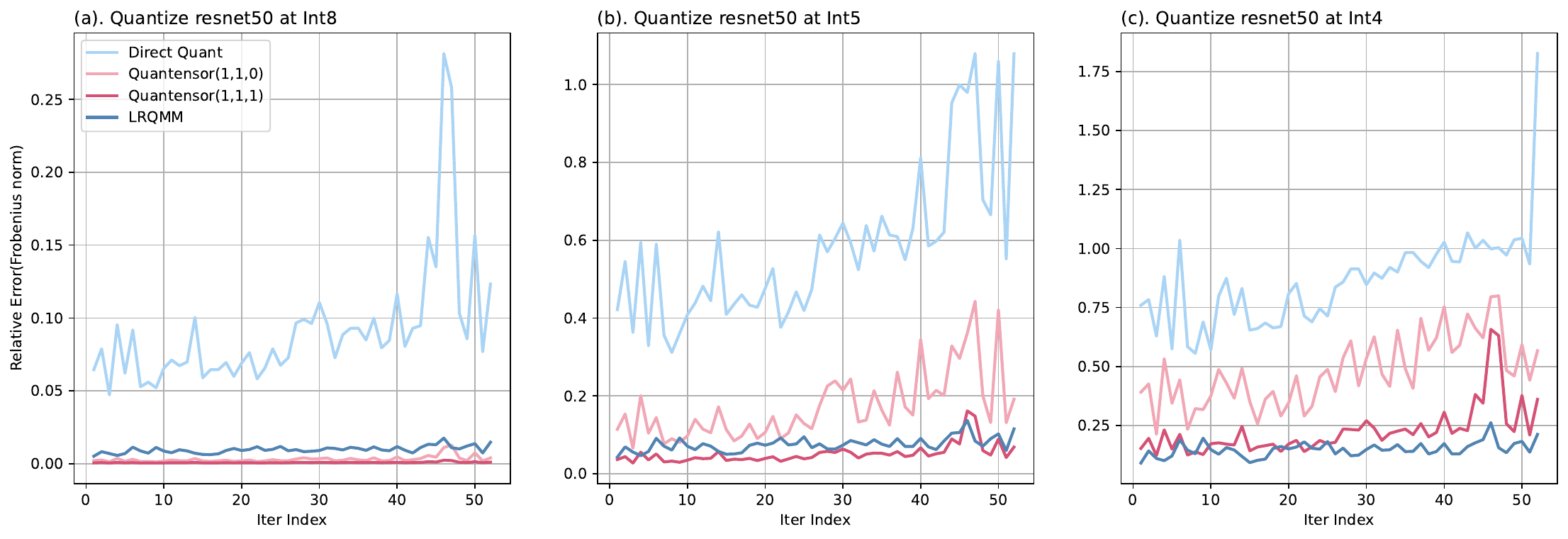}
	\caption{
 In deep learning networks, the relative error of different quantization algorithms at each layer.
}
	\label{errnet}
\end{figure*}

\begin{table}[htbp]
  \centering
  \caption{Comparison with state-of-the-art quantization methods on ImageNet. And \textit{QT} is short for QuantTensor. \textit{Direct Quant} uses the first term in Eq.\ref{gemm_r_split}, and QT(1,1,0), QT(1,1,1) use the first three and full four terms in Eq.\ref{gemm_r_split}}.
    \begin{tabular}{ccccc}
    \toprule
    \multirow{2}[2]{*}{Methods} & \multicolumn{4}{c}{Accuracy at Bit-width} \\
          & \multicolumn{4}{c}{(W/A)} \\
    \midrule
    \midrule
    \multicolumn{5}{c}{Resnet-18 (FP: 70.2)} \\
          & 8/8   & 5/5   & 4/4   & 3/3 \\
    \hdashline
    QT(1,1,0)\cite{li2021unleashing} & 69.8  & 62.5  & 27.7  & 8.81 \\
    QT(1,1,1)\cite{li2021unleashing} & 71.9  & 69.6  & 62.2  & 57.1 \\
    AdaDFQ\cite{ADADFQ} & \textbf{-} & 70.2  & 66.5  & 38.1 \\
    AdaSG\cite{ADASG} & \textbf{-} & 70.3  & 66.5  & 37.0 \\
    IntraQ\cite{Intraq} & \textbf{-} & 66.74 & 66.74 & \textbf{-} \\
    ABC-Net\cite{ABCNET} & \textbf{-} & 65.0  & \textbf{-} & 61.0 \\
    Direct Quant & 60.2  & 28.5  & 14.2  & 0.33 \\
    \textbf{LRQMM(ours)} & 69.4  & 64.7  & 58.5  & 39.5 \\
    \midrule
    \midrule
    \multicolumn{5}{c}{Resnet-34 (FP: 73.3)} \\
          & 8/8   & 5/5   & 4/4   & 3/3 \\
    \hdashline
    QT(1,1,0)\cite{li2021unleashing} & 72.7  & 66.2  & 30.9  & 7.71 \\
    QT(1,1,1)\cite{li2021unleashing} & 72.8  & 70.1  & 67.1  & 57.7 \\
    AdaDFQ\cite{ADADFQ} & \textbf{-} & \textbf{-} & \textbf{-} & \textbf{-} \\
    AdaSG\cite{ADASG} & \textbf{-} & \textbf{-} & \textbf{-} & \textbf{-} \\
    IntraQ\cite{Intraq} & \textbf{-} & \textbf{-} & \textbf{-} & \textbf{-} \\
    ABC-Net\cite{ABCNET} & \textbf{-} & 68.4  & \textbf{-} & 66.7 \\
    Direct Quant & 72.5  & 20.2  & 11.1  & 0.32 \\
    \textbf{LRQMM(ours)} & 72.7  & 68.4  & 60.2  & 39.6 \\
    \midrule
    \midrule
    \multicolumn{5}{c}{Resnet-50 (FP: 76.0)} \\
          & 8/8   & 5/5   & 4/4   & 3/3 \\
    \hdashline
    QT(1,1,0)\cite{li2021unleashing} & 75.7  & 64.2  & 30.5  & 8.32 \\
    QT(1,1,1)\cite{li2021unleashing} & 75.8  & 73.4  & 68.9  & 60.5 \\
    AdaDFQ\cite{ADADFQ} & \textbf{-} & 76.1  & 68.4  & 17.6 \\
    AdaSG\cite{ADASG} & \textbf{-} & 76.0  & 68.6  & 16.9 \\
    IntraQ\cite{Intraq} & \textbf{-} & \textbf{-} & \textbf{-} & \textbf{-} \\
    ABC-Net\cite{ABCNET} & \textbf{-} & 70.1  & \textbf{-} & \textbf{-} \\
    Direct Quant & 75.2  & 15.2  & 8.3   & 0.33 \\
    \textbf{LRQMM(ours)} & 75.6  & 72.4  & 61.8  & 41.9 \\
    \bottomrule
    \end{tabular}%
  \label{resnet18}%
\end{table}%

\textbf{Differential distribution:}
The Table.\ref{LRQMMM_int4} and Table.\ref{LRQMMM_int8} demonstrate the accuracy of the algorithm under int4 and int8 quantization for different distributions, with a matrix size of 2000 and an approximate rank of 10.

From these two tables, it can be observed that the algorithm has a significant effect on the residual compensation of low-precision computations. With int4, even in the worst-case scenario of normal distribution, there is nearly \textbf{1} order of magnitude improvement in accuracy compared to the direct quantization method, which is already comparable to Quantensor(1,1,1). In the best-case scenarios of uniform and exponential distributions, there is \textbf{2\~3} orders of magnitude improvement improvement in accuracy. In the case of int8, LRQMM has an improvement in accuracy of one or two orders of magnitude. Particularly, the algorithm exhibits exceptionally excellent performance in exponential and chi-square distributions whereas direct quantization methods perform poorly.

It is particularly worth mentioning that the algorithm uses an approximate rank of only 10. In the case of larger matrix scales, the algorithm only requires a time complexity increase at the level of \( N^2 \), which is very minimal compared to the \( N^3 \) time complexity of matrix multiplication. This can also be seen from the subsequent algorithmic time proportion tests.

\textbf{Image recognition:} The convolutional operations were transformed into matrix multiplications using the img2col method, followed by training the ResNet architecture network on the ImageNet dataset with LRQMM(rank=20). The results, which are presented in Table.\ref{resnet18} for different quantization bit-widths along with the relative error of different quantization algorithms at each layer presented in Fig.\ref{errnet}.

It is worth mentioning that only direct quantization algorithms were used in the implementation of LRQMM, without employing other methods such as KL divergence or quantization parameter learning. Therefore, in some datasets, it has not surpass the state-of-the-art.

By adding the calculation of low-rank residual terms, the overall accuracy can be significantly improved. From the results, LRQMM achieving accuracy close to that of Quantensor(1,1,1) methods at lower precision levels, demonstrate that LRQMM offers significantly better accuracy compared to direct quantization and two improved methods, and exhibits an even greater advantage in low-bitwidth networks (below 4 bits).

\subsection{Performance of LRQMM}

\begin{figure}
	\centering
\includegraphics[width=1\linewidth]{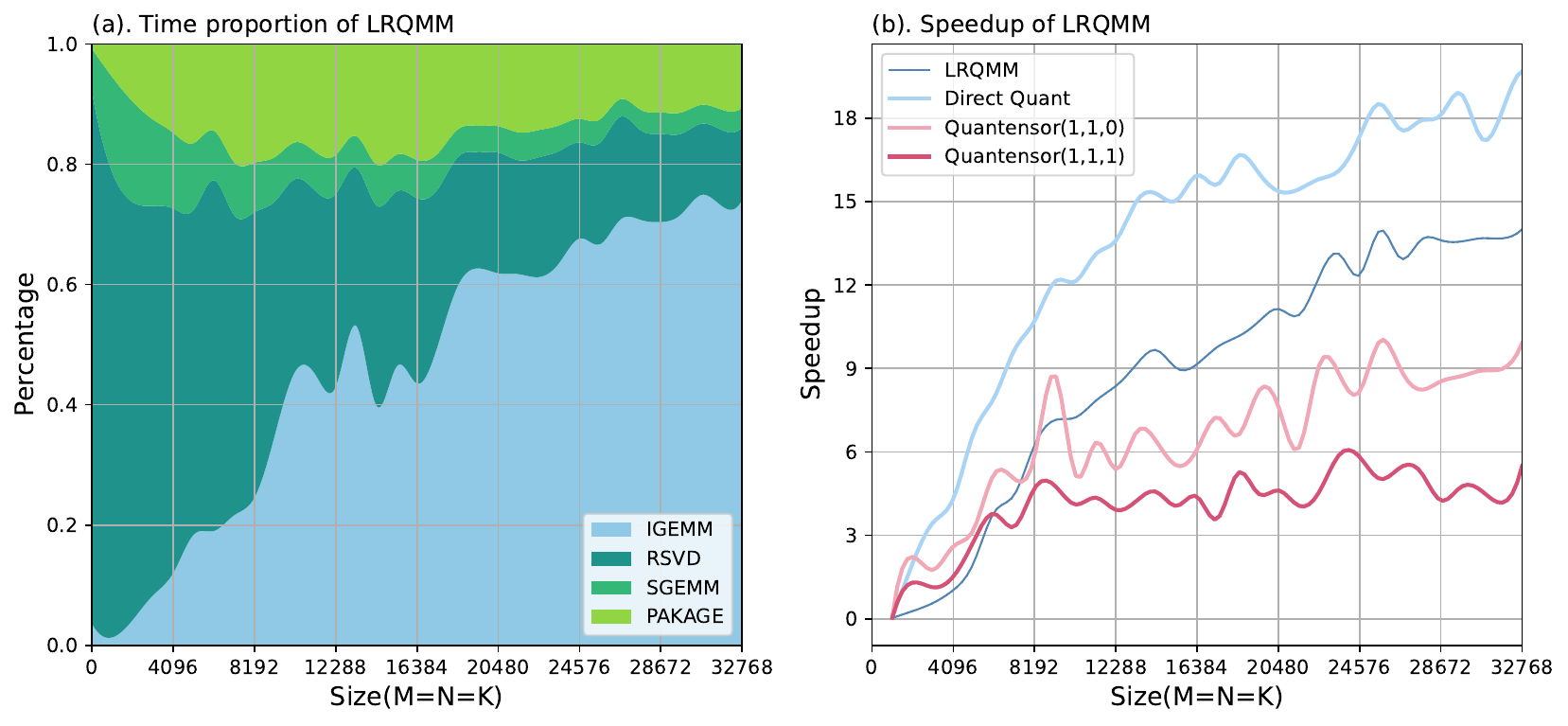}
	\caption{
 (a). Time proportion of different parts of the algorithm, where \textit{PAKAGE} accounts for the time needed for matrix addition, quantization, and other operations aside from the aforementioned three items. (b). Different quantization methods speedup on the GPU, and the baseline is \textit{SGEMM} provided by cuBLAS.
}
	\label{Speedup_propotion}
\end{figure}

\textbf{Algorithm Proportion:}
Due to the accuracy tests in the previous section having demonstrated the stability of the algorithm. In this part, we utilize an approximate rank of 10. The compositional time proportion of the algorithm in GPU is illustrated in Fig.\ref{Speedup_propotion} (a).

It is noteworthy that the the RSVD algorithm in cuSolver is only available in single precision and the performance is not optimal when the matrix is small in size. As the matrix size increases, the execution time of the algorithm reverts to being primarily composed of low-precision matrix multiplication, aligning with our estimates for the algorithm. In summary, with good algorithm implementations, LRQMM demonstrates considerable performance.

\textbf{Algorithm Speedup:}
The LRQMM was tested alongside several comparative algorithms on the GPU platform, with the resulting speedup ratios depicted in Figure.\ref{Speedup_propotion}. (b). The algorithm exhibits a distinct advantage when dealing with larger matrix scales, with speedup ratios approaching those of direct quantization methods.

By combining this figure with Table 3, it can be observed that compared to the QT algorithm, LRQMM can achieve over 40\% performance improvement at larger matrix scales while enhancing the algorithmic accuracy more than 90\% in most distributions (Uniform(0,1) , Exponent, ChiSquare, Poisson).

\section{Discussion}
\quad \textbf{Generality and Potential Applications: } We have demonstrated the effectiveness of LRQMM in accelerating GEMM and neural networks. Our experimental results indicate that LRQMM has potential efficacy for many other tensor programs in practical applications. Due to the diversity of tensor programs, the acceleration and precision loss achieved by LRQMM will vary depending on the application. However, the main components of the LRQMM algorithm are loosely coupled, allowing developers to optimize their programs by trading off performance and precision specific to their applications. 

\textbf{Quantization methods:} Currently, the LRQMM algorithm only employs direct online symmetric quantization methods, as they are relatively easy to implement and do not incur additional storage parameter overhead. In current deep learning applications, research on quantization algorithms, such as learning-based quantization parameters, KL divergence, and non-symmetric quantization\cite{chen2019metaquant}, can also enhance the precision of deep learning applications. Moreover, LRQMM can be easily integrated with these methods to further improve the accuracy of quantization.

\section{Conclusion}
This paper presents LRQMM algorithm, which is a data free operator level quantization optimization method. LRQMM combines residual compensation quantization with random low-rank decomposition technology, avoids the calculation of the whole matrix in the process of residual calculation, uses the low-rank property of residual matrix to improve the accuracy of the overall quantization algorithm, and is easy to combine with other quantization methods. It is proved that the accuracy of this algorithm in deep learning applications are significantly improved. In future work, we plan to improve the applicability of RSVD in this method to improve the overall efficiency of the algorithm and explore more application scenarios for LRQMM.

\begin{acks}                            
  This material is based upon work supported by the
  \grantsponsor{GS100000001}{National Science
    Foundation}{http://dx.doi.org/10.13039/100000001} under Grant
  No.~\grantnum{GS100000001}{nnnnnnn} and Grant
  No.~\grantnum{GS100000001}{mmmmmmm}.  Any opinions, findings, and
  conclusions or recommendations expressed in this material are those
  of the author and do not necessarily reflect the views of the
  National Science Foundation.
\end{acks}

\clearpage

\clearpage

\end{document}